# Hybrid Molecular Beam Epitaxy of Ge-based Oxides

Fengdeng Liu[1,†], Tristan K Truttmann[1,†], Dooyong Lee[1], Bethany E. Matthews[4], Iflah Laraib[2], Anderson Janotti[2], Steven R. Spurgeon[4,5], Scott A. Chambers[3], and Bharat Jalan[1,*]

[1] Department of Chemical Engineering and Materials Science, University of Minnesota, Minneapolis, Minnesota 55455, USA

[2] Department of Materials Science and Engineering, University of Delaware, Newark, Delaware 19716, USA

[3] Physical and Computational Sciences Directorate, Pacific Northwest National Laboratory, Richland, Washington 99352, USA

[4] Energy and Environment Directorate, Pacific Northwest National Laboratory, Richland, Washington 99352, USA

[5] Department of Physics, University of Washington, Seattle, Washington 98195, USA

† FL and TKT are equally contributing authors.

* All correspondence should be addressed to BJ (bjalan@umn.edu).

**Abstract**

Germanium-based oxides such as rutile $GeO_2$ are garnering attention owing to their wide band gaps and the prospects for ambipolar doping for application in high-power devices. Here, we present the use of germanium tetraisopropoxide (GTIP) (an organometallic chemical precursor) as a source of Ge for the demonstration of hybrid molecular beam epitaxy (MBE) for Ge-containing compounds. We use $Sn_{1-x}Ge_xO_2$ and $SrSn_{1-x}Ge_xO_3$ as model systems to demonstrate this new synthesis method. A combination of high-resolution X-ray diffraction, scanning transmission electron microscopy, and X-ray photoelectron spectroscopy confirms the successful growth of epitaxial rutile $Sn_{1-x}Ge_xO_2$ on $TiO_2$(001) substrates up to $x = 0.54$ and coherent perovskite $SrSn_{1-x}Ge_xO_3$ on $GdScO_3$(110) substrates up to $x = 0.16$. Characterization and first-principles calculations corroborate that Ge preferentially occupies the Sn site, as opposed to the Sr site. These findings confirm the viability of the GTIP precursor for the growth of germanium-containing oxides by hybrid MBE, and thus open the door to high-quality perovskite germanate films.

**Introduction**

The holy grail of semiconductor heterostructures is the ability to produce high-quality semiconductor films with tunable bandgaps that are also lattice-matched to commercially available substrates. One common strategy to achieve this is to alloy two or more semiconductors and judiciously choose the alloy composition for the desired bandgap and lattice parameter. In principle, this approach can be applied to the alkaline earth stannates, a system of three materials (namely $CaSnO_3$, $SrSnO_3$, and $BaSnO_3$) that has recently fascinated researchers for potential applications in next-generation power electronics and ultraviolet optoelectronics. The alloys of these three materials, whose crystal structures are summarized in Figure 1(b), span the purple shaded region of bandgaps ($E_g$) and pseudocubic lattice parameters ($a$) in Figure 1(a). However, the scarcity of commercially available substrates makes the range of accessible bandgaps in lattice-matched systems rather limited. The use of $GdScO_3$ as the substrate facilitates the largest bandgap range for lattice-matched alloys. However, the span is only 0.1 eV, from $E_g$ = 4.2 eV ($Ba_{0.13}Ca_{0.87}SnO_3$) to $E_g$ = 4.3 eV ($Sr_{0.26}Ca_{0.74}SnO_3$). Such a limited bandgap range provides few opportunities for modulation doping using lattice-matched oxide heterostructures.

Adding a second tuning parameter to this material system would expand the gamut of accessible properties. Replacement of Sn with Ge offers one such tuning parameter. Oxides containing $Ge^{4+}$ typically have conduction bands derived from Ge 4$s$ orbitals, which, analogous to Sn 5$s$ orbitals, produce dispersive conduction bands with low electron effective masses. However, the different sizes and energies of the atomic orbitals lead to nuanced differences in the physical and electronic structures. This variation has been exploited in rutile oxides, where the substitution of Sn with Ge yields semiconductors with bandgaps ranging from 3.6 eV to 4.7 eV [1] with predicted ambipolar doping, offering encouraging prospects as ultrawide bandgap (UWBG) semiconductors for power electronics [2].

The crystal structures of alkaline earth germanates are summarized in Figure 1(c). While perovskite BaGeO$_3$—a chemical analog to BaSnO$_3$—has not yet been synthesized, it is predicted to be metastable in the cubic structure [3]. The cubic perovskite SrGeO$_3$ has been successfully synthesized by a high-pressure, high-temperature method and quenched to ambient conditions; it displays an indirect band gap $E_g$ = 2.7 eV, but its wider *direct* band gap (3.5 eV) makes it transparent to visible light [4]. SrGeO$_3$ has also been doped with La, yielding a room-temperature mobility of 12 cm$^2$V$^{-1}$s$^{-1}$ [4]. However, DFT calculations suggest that both SrGeO$_3$ and BaGeO$_3$ have the potential to achieve phonon-limited mobilities superior to those of BaSnO$_3$ [3]. Finally, CaGeO$_3$ has a metastable orthorhombic perovskite structure quenchable to ambient conditions. Although this material has not been optically or electrically characterized, it is almost certainly a transparent semiconductor with $E_g$ > 2.7 eV [5-8].

The yellow region in Figure 1(a) shows the additional range of $E_g$ and $a$ made available by adding SrGeO$_3$ to the stannate alloy system. We note that the region encompasses nearly all commercially available perovskite oxide substrates, and alloys lattice-matched to DyScO$_3$ substrates have band gaps that span 1.5 eV, from $E_g$ = 2.9 eV (Ba$_{0.49}$Sr$_{0.51}$Sn$_{0.49}$Ge$_{0.51}$O$_3$) to $E_g$ = 4.4 eV (Sr$_{0.07}$Ca$_{0.93}$SnO$_3$), providing ample opportunity for lattice-matched heterostructures in modulation-doping field-effect transistors (MODFET) and, potentially, even optoelectronic devices.

Molecular beam epitaxy (MBE) is considered a gold-standard technique to grow high-quality heterostructures. However, since MBE relies on the codeposition or shutter growth of individual elements, achieving a composition with a target $E_g$ and $a$ while simultaneously maintaining the A:B-site cation stoichiometry presents a formidable challenge in flux calibration. Therefore, it is desirable to use adsorption-controlled growth, which exploits thermodynamics to automatically regulate the A:B-site cation stoichiometry. Hybrid MBE is a technique that draws on the high vapor pressure of metal-organic precursors to achieve adsorption-controlled growth. It has been successfully applied to the adsorption-controlled growth of titanates [9], vanadates [10], and stannates [11]. However, it has not yet been applied to the growth of Ge-containing oxides.

Here, we report on a hybrid MBE approach for the growth of Ge-based oxides using germanium tetraisopropoxide (GTIP) as a metal-organic precursor for Ge. Figure 2 shows that GTIP has a much higher vapor pressure than elemental Ge and is comparable to established metal-organic precursors for hybrid MBE. We chose $Sn_{1-x}Ge_xO_2$ and $SrSn_{1-x}Ge_xO_3$ as model systems to demonstrate the use of GTIP in the growth of binary and ternary oxides, successfully synthesizing epitaxial rutile $Sn_{1-x}Ge_xO_2$ and coherent perovskite $SrSn_{1-x}Ge_xO_3$ films.

## Methods
### Film Growth

$Sn_{1-x}Ge_xO_2$ ($x$ = 0, 0.28, 0.54) and $SrSn_{1-x}Ge_xO_3$ ($x$ = 0, 0.05, 0.08, 0.16) films were successfully grown using hybrid MBE. This approach employs a conventional solid source for Sr where necessary, hexamethylditin (HMDT) as a metal-organic precursor for Sn, germanium tetraisopropoxide (GTIP) as a metal-organic precursor for Ge, and an inductively coupled RF plasma for O. Rutile films were grown on $TiO_2$(001) substrates at 600 °C, and perovskite films were grown on $GdScO_3$(110) substrates at 950 °C. When Sr was used, its beam equivalent pressure (BEP)—measured by a retractable beam flux monitor—was fixed at $2.3 \times 10^{-8}$ Torr. The BEP ratio of HMDT and GTIP was varied to control the Sn:Ge ratio in the films. The oxygen flow was set to 0.7 sccm to achieve an oxygen background pressure of $5 \times 10^{-6}$ Torr while applying 250 W RF power to the plasma coil. Each rutile film was grown for 60 minutes, and the perovskite films were grown for 30 minutes.

**Film Characterization**

Surface topography was measured with a Bruker Nanoscope V Multimode 8 atomic force microscope (AFM). A Rigaku SmartLab XE was used for X-ray scattering. High-resolution X-ray diffraction (HRXRD) 2*q*-*w* coupled scans were used to determine out-of-plane lattice parameters. Film thickness was determined from HRXRD finite thickness fringes when possible, or otherwise extracted from grazing incidence X-ray reflectivity (GIXR). Reciprocal space maps (RSMs) were used to measure in-plane lattice parameters and determine strain relaxation. On each RSM, Vegard's law was used to calculate relaxed lattice parameters and Poisson ratios were used to calculate how the lattice parameters would change under biaxial stress. For the rutile system, lattice parameters were taken from powder neutron diffraction [12,13], and Poisson ratios were interpolated from elastic tensor data of the end members [14,15]. For the perovskite system, substrate lattice parameters from Liferovich and coworkers [16] were used, whereas the $SrSnO_3$ tetragonal lattice parameters from Glerup and coworkers [17] and the parameters of ambient temperature (quenched) $SrGeO_3$ from Nakatsuka [18] were used for the film. Due to a lack of elastic tensor data for $SrGeO_3$, the DFT-predicted Poisson ratio for $SrSnO_3$ of 0.192 [19] was used for the entire alloy series.

X-ray photoelectron spectroscopy (XPS) was used to determine the Ge concentration and valence. To determine composition, survey scans were measured using a Physical Electronics 5000 VersaProbe III photoelectron spectrometer with monochromatic Al Kα x-rays at the University of Minnesota. To determine Ge valence, XPS was also performed at PNNL using an Omicron/Scienta R3000 analyzer with monochromatic Al Kα x-rays, a 100 eV analyzer pass energy, a 0.8 mm slit width, and a normal emission geometry. The resulting energy resolution was ~400 meV as judged by fitting the Fermi edge for a clean, polycrystalline Ag foil to the Fermi-Dirac function. The insulating nature of these films required the use of a low-energy electron flood gun to compensate the positive photoemission charge that builds up on the surface. The flood gun makes it possible to measure accurate core-level (CL) line shapes on insulating samples. However, the measured binding energies are artificially low due to overcompensation. At the same time, we sought to use a highly conductive *n*-Ge(001) crystal with its thin native oxide as an internal binding energy standard for $Ge^0$, $Ge^{2+}$, $Ge^{3+}$ and $Ge^{4+}$, as assigned by Molle and coworkers [20]. In order to compare Ge 3*d* binding energies from the insulating MBE-grown films to those from the $GeO_x$/Ge standard, all samples were affixed to an insulating MgO(001) wafer in order to uniformly isolate them from ground. The flood-gun beam energy was set to ~1 eV. The charging-induced binding energy shifts were close, but not identical, for the $GeO_x$/*n*-Ge sample and the epitaxial films, as judged by the aliphatic C 1s peak binding energy from the surface contamination. After correcting for differences in charging as judged by the C 1s binding energies, all spectra were shifted by a constant amount such that the Ge $3d_{5/2}$ lattice peak in the $GeO_2$/*n*-Ge spectrum fell at 29.4 eV, the value measured when $GeO_2$/*n*-Ge(001) is mounted directly on the grounded sample holder without an MgO wafer for electrical isolation.

Cross-sectional scanning transmission electron microscopy (STEM) samples were prepared using a FEI Helios NanoLab DualBeam $Ga^+$ Focused Ion Beam (FIB) microscope with a standard lift out procedure. STEM high-angle annular dark field (STEM-HAADF) images were collected on a probe-corrected JEOL GrandARM-300F microscope operating at 300 kV, with convergence semi-angle of 29.7 mrad and a collection angle range of 75–515 mrad. For the STEM-HAADF images shown in Figure 5(c), a series of 10 frames was collected at 512 × 512 px sampling, with a dwell time of 2 μs $px^{-1}$. The images were rigid-aligned using the SmartAlign program to minimize drift and then averaged to improve signal-to-noise [21]. Simultaneous STEM energy dispersive X-ray spectroscopy (STEM-EDS) and electron energy-loss spectroscopy (STEM-EELS) elemental mapping was used to determine site occupation, using the Sn $L$ peak and Sr $L$ and Ge $L$ edges, respectively. For this configuration, a STEM-EELS acceptance angle range of 113–273 mrad was used. Mapping was performed using a dual JEOL Centurio detector setup for STEM-EDS and a 1 eV ch$^{-1}$ dispersion, with a 4× energy binning in the dispersive direction for STEM-EELS. No denoising was applied, but the composite map shown in Figure 5(d) was filtered using a smoothing kernel in Gatan Microscopy Suite 3.4.3.

**First-Principles Calculations**

First-principles calculations were performed to investigate whether Ge prefers the octahedrally coordinated Sn site with oxidation state 4+, or the Sr site with oxidation state 2+. The calculations are based on density functional theory [22,23] within the generalized gradient approximation revised for solids (PBEsol [24]), with projector augmented wave potentials [25,26] as implemented in the VASP code [27,28]. We considered both $SrSn_{1-x}Ge_xO_3$ and $Sr_{1-x}Ge_xSrO_3$ using supercells containing 20, 40, and 80 atoms representing concentrations $x = 0.0625$, 0.125, and 0.25. We used an energy cutoff of 500 eV for the plane wave expansions and meshes of $k$-points that are equivalent to 6×6×4 for the 20-atom cell of the tetragonal $SrSnO_3$. All the atom positions in the cell were allowed to relax. To simulate the epitaxial growth of the alloy on $GdScO_3$(110) substrates, we fixed the in-plane lattice parameters to that of $GdScO_3$, allowing the out-of-plane lattice parameter to relax.

**Results and Discussion**

Figure 3(a) shows AFM of rutile $Sn_{1-x}Ge_xO_2/TiO_2$(001) with different Ge concentrations. Increasing the germanium fraction $x$ from 0 to 0.54 decreased the root mean square (RMS) roughness from 1373 pm to 461 pm. These RMS roughnesses are roughly proportional to the thickness of the films, which decrease with Ge incorporation. As explained in Supplementary Note 1, the thickness is not a simple function of the precursor BEP, and the variation is best explained by the desorption of Ge- and Sn-containing species.

Figure 3(b) shows the rutile HRXRD $2\theta$-$\omega$ coupled scans and corresponding rocking curves around the (002) film peaks. The $2\theta$-$\omega$ coupled scans show that the film lattice parameters decrease with Ge incorporation, consistent with Ge's smaller ionic radius. The full width at half maximum (FWHM) of each film *decreases* from 0.93° to 0.086° as Ge incorporation increases from 0 to 0.54. This goes against the intuitive expectation that adding Ge would increase the structural disorder by disrupting translational symmetry through random substitution. It is also noteworthy that these rocking curves appear to be a linear combination of two Gaussians (a narrow and a broad component). We explain both observations in the discussion of the RSMs below in terms of strain relaxation.

Additionally, we applied GTIP-based hybrid MBE to Ge-based ternary oxides. Figure 3(c) shows AFM images of $SrSn_{1-x}Ge_xO_3$/$GdScO_3$(110) with different Ge concentrations. These micrographs show surface roughnesses that decrease with Ge incorporation from 503 pm to 171 pm. Unlike the rutile samples, however, this trend cannot be explained by film thickness. Instead, the perovskite film thickness is not affected by Ge incorporation, suggesting the Sn-species desorption is not affected by Ge incorporation in this material system. The decreased surface roughness with Ge incorporation might instead be explained by other factors, such as modified adatom mobility or a decreased driving force for adatom agglomeration.

Figure 3(d) shows the HRXRD $2\theta$-$\omega$ coupled scans and corresponding rocking curves around the perovskite $(002)_{pc}$ film peak. The $2\theta$-$\omega$ coupled scans demonstrate that the replacement of Sn with Ge decreases the film lattice parameters, consistent with smaller $Ge^{4+}$ at the $Sn^{4+}$ site. The Kiessig fringes and rocking curve FWHM of 0.07-0.08° demonstrate uniform films with high structural quality.

To investigate how Ge-incorporation influences the strain relaxation of these films, we measured reciprocal space maps of both the rutile and perovskite samples. Figure 4(a-c) shows the RSMs around the (202) reflection of the rutile films. Contours and guidelines have been added to show the expected peak position depending on composition and strain, following a similar procedure used for (Al,Ga)N heterostructures by Enslin and coworkers [29]. Each contour represents all possible strains at a single composition, and the two guidelines represent all possible compositions for the fully coherent and fully relaxed films. For the $SnO_2$ film in Figure 4(a), the film peak is centered over the $x = 0$ contour, close to where it intersects the relaxed guideline suggesting a nearly complete film relaxation.

For the film in Figure 4(b), the film peak resides slightly north of the $x = 0.25$ contour, consistent with the $x = 0.28$ Ge fraction determined from XPS. Furthermore, the peak lies between the relaxed and coherent guidelines, indicating the film is compressively strained and has undergone a small degree of relaxation toward its bulk lattice parameter. For the $x = 0.54$ film in Figure 4(c), the film peak resides directly over the coherent guideline, indicating a film that is completely tensile strained to the substrate. The reader may notice that the film is expected to lie north of the $x = 0.50$ contour but is in fact south of it. This small discrepancy is mostly likely due to small deviations from Vegard's law, which was used to calculate the positions of these contours. This discrepancy may also be caused by error in the composition determined from XPS.

The progression from a nearly fully relaxed film at $x = 0$ to a fully strained film at $x = 0.54$ can be explained by considering two facts. First, the growth rates decrease with Ge incorporation, so higher values of *x* correspond to thinner films which have less built-up elastic strain energy. Second, films with higher Ge fractions have smaller lattice parameters, better matching the $TiO_2$ substrate, also resulting in less elastic strain energy. The resulting trend in strain relaxation fully explains why Ge incorporation improves the rocking curves in Figure 3(b). The two-Gaussian shaped rocking curve is a well-understood phenomenon commonly observed during the strain relaxation of epitaxial films [30].

The RSMs of three representative *perovskite* films are shown in Figures 4(d-f). Unlike the rutile films, each perovskite film is fully strained to the $GdScO_3$ substrate. Again, one may notice small deviations between the film peak positions and their expected position based on the composition contours. For example, the *x* = 0 film peak in Figure 4(d) is slightly north of the *x* = 0 contour. These discrepancies are probably due to a lack of accurate experimental Poisson ratios used to calculate the contours.

One major challenge associated with growth of high-quality oxides containing late transition metals (like Ru, Ni, and Cu) or main group metals (like Bi, Ge, and Sn) is achieving full oxidation of these high-electronegativity metals. To investigate the oxidation of Ge, we performed XPS on the rutile and perovskite films. Figure 5(a) shows Ge 3*d* core-level spectra of *rutile* films compared to that of a Ge reference wafer with native oxide; we mark the Ge 3*d* binding energy of different valence states for comparison, as shown at the top of Figures 5(a) and 5(b) using assignments from Molle and coworkers [20]. We can clearly see that film peak position matches the $Ge^{4+}$ position in the reference wafer, suggesting that Ge stays in the 4+ state in which it is delivered via GTIP.

In the XPS of the *perovskite* samples shown in Figure 5(b), however, the film peak position better matches the $Ge^{3+}$ position in the reference wafer, which suggests that Ge is in the 3+ state.

However, this is unlikely because Ge is generally not stable in the 3+ state. The Ge in the reference wafer was oxidized by exposure to air where oxidation is limited by solid-state diffusion. Hence, the reference wafer can achieve the otherwise unattainable oxidation states of 1+ and 3+, which still only constitute a very small fraction of the analyzed volume. We posit that the unusual coordination environment for Ge in the perovskite structure results in a different binding energy for B-site $Ge^{4+}$ than is observed in amorphous $GeO_2$ due to differences in total electrostatic potential.

To determine the coordination environment of the Ge guest in the $SrSnO_3$ host lattice, cross-sectional STEM was performed. Figure 5(c) shows STEM-HAADF images of two perovskite samples, in which image intensity scales with atomic number ($Z^{~1.7}$). The images show high quality interfaces free of dislocations, consistent with the conclusion of fully coherent films determined from RSMs in Figure 4(e-f). The $x = 0.08$ sample shows a subtle low-$Z$ band at the interface. This feature might be a result of slight A:B-site nonstoichiometry due to effusion cell temperature transients caused by opening the shutter at the beginning of growth. Figure 5(d) shows atomic-resolution STEM-EDS and STEM-EELS elemental maps and line profiles of Ge, Sr, and Sn. The line profiles show clear alignment of Ge and Sn peaks occurring in the valleys of the Sr signal, demonstrating direct substitution of Sn with Ge on the B-site. Therefore, our experimental data indicate that Ge resides in an octahedral coordination environment in the perovskite lattice.

To further examine the site preference of Ge, we have carried out DFT calculations to determine the formation enthalpy of $Sr_{1-x}Ge_xSnO_3$ (referred to as A-site alloys) and $SrSn_{1-x}Ge_xO_3$ (referred to as B-site alloys) as function of Sn, Ge, and Sr chemical potentials. We can define the preference of Ge occupying Sr- vs. Sn-site by taking the difference in the formation enthalpies of A-site and B-site alloys as function of Sn and O chemical potentials, such as:

$$\Delta H^f [\text{Sr}_{1-x}\text{Ge}_x\text{SnO}_3] =$$

$$E_t[\text{Sr}_{1-x}\text{Ge}_x\text{SnO}_3] - E_t(\text{SrSnO}_3) + [(1-x)E_t(\text{Sr}) + \mu_{\text{Sr}}] - x[E_t(\text{Ge}) + \mu_{\text{Ge}}] \quad (1)$$

and

$$\Delta H^f [\text{SrSn}_{1-x}\text{Ge}_x\text{O}_3] =$$

$$E_t[\text{SrSn}_{1-x}\text{Ge}_x\text{O}_3] - E_t(\text{SrSnO}_3) + [(1-x)E_t(\text{Sn}) + \mu_{\text{Sn}}] - x[E_t(\text{Ge}) + \mu_{\text{Ge}}] \quad (2)$$

where $E_t[\text{Sr}_{1-x}\text{Ge}_x\text{SnO}_3]$ and $E_t[\text{SrSn}_{1-x}\text{Ge}_x\text{O}_3]$ are total energies of the alloys, $E_t(\text{SrSnO}_3)$ is the total energy of the host material, and $E_t(\text{Sr})$, $E_t(\text{Sn})$, and $E_t(\text{Ge})$, and are total energy per atom of the Sr, Sn, and Ge bulk phases, to which the chemical potentials $\mu_{\text{Sr}}$, $\mu_{\text{Sn}}$, and $\mu_{\text{Ge}}$ are referenced ($\mu_{\text{Sr}}, \mu_{\text{Sn}}, \mu_{\text{Ge}} \leq 0$). These chemical potentials are not independent, but must satisfy the stability condition of the parent material SrSnO3, i.e.,

$$\mu_{\text{Sr}} + \mu_{\text{Sn}} + 3\mu_{\text{O}} = \Delta H^f(\text{SrSnO}_3) \quad (3)$$

avoiding the formation of secondary phases SrO, SnO2, and GeO2, i.e.,

$$\mu_{\text{Sr}} + \mu_{\text{O}} < \Delta H^f(\text{SrO}), \quad (4)$$

$$\mu_{\text{Sn}} + 2\mu_{\text{O}} < \Delta H^f(\text{SnO}_2), \quad (5)$$

and

$$\mu_{\text{Ge}} + 2\mu_{\text{O}} < \Delta H^f(\text{GeO}_2). \quad (6)$$

Combining Eqs. (3) and (4), we obtain $\mu_{\text{Sn}} + 2\mu_{\text{O}} > \Delta H^f(\text{SrSnO}_3) - \Delta H^f(\text{SrO})$, which together with Eqs. (5) and (6) limit the region in the $\mu_{\text{Sn}}$ vs. $\mu_{\text{O}}$ diagram where SrSnO3 is stable, as shown in Figure 6. The lines separating the region where the B-site alloys $\text{SrSn}_{1-x}\text{Ge}_x\text{O}_3$ have lower formation enthalpies than the A-site alloys $\text{Sr}_{1-x}\text{Ge}_x\text{SnO}_3$, i.e., $\Delta H^f[\text{SrSn}_{1-x}\text{Ge}_x\text{O}_3] \leq \Delta H^f[\text{Sr}_{1-x}\text{Ge}_x\text{SnO}_3]$, are located in the upper right corner of the $\mu_{\text{Sn}}$ vs $\mu_{\text{O}}$ diagram.

We can see that SrSnO$_3$ and SrSn$_{1-x}$Ge$_x$O$_3$ alloys are only stable in the orange region at the center of Figure 6, limited by the formation of SrO (left), GeO$_2$ (below) and SnO$_2$ (right). The conditions for which the formation enthalpies of the A-site alloys Sr$_{1-x}$Ge$_x$SnO$_3$ would be lower than that of B-site alloys SrSn$_{1-x}$Ge$_x$O$_3$ occur in a region of $\mu_{Sn}$ and $\mu_O$ chemical potentials where the secondary phase SnO$_2$ is most favorable to form, i.e., where SrSnO$_3$ itself is unstable. This result clearly indicates that Ge prefers the Sn octahedral site over the Sr site.

We also note that in the A-site alloys, the Ge atom displays a large offsite displacement toward 3 of the original 12 nearest-neighbor O atoms (along the [110] pseudocubic direction), with Ge-O distances of ~2.06 Å. This further indicates that Ge$^{2+}$ on the Sr site is unstable. In contrast, Ge sits on the Sn octahedral sites in SrSn$_{1-x}$Ge$_x$O$_3$, with Ge-O distances of ~1.93 Å (equatorial) and of ~1.97 Å (apical), reflecting the tetragonal structure, and shows no offsite displacement. These results, again, reflect the fact that Ge strongly prefers the octahedral Sn site over the Sr site in SrSnO$_3$, consistent with our experimental observations.

**Summary**

In summary, we have demonstrated the growth of epitaxial Sn$_{1-x}$Ge$_x$O$_2$ and SrSn$_{1-x}$Ge$_x$O$_3$ films via hybrid MBE. AFM, HR-XRD, XPS, and STEM characterization shows that the GTIP precursor can be used as an effective source of Ge for the growth of both rutile and perovskite oxides while allowing excellent surface morphology and structural quality. DFT calculations indicate that Ge strongly prefers the Sn site in SrSnO$_3$ as opposed to the Sr site. This work opens another synthetic route to achieving Ge-containing oxides. Future studies should build upon this work by exploring process parameters to achieve phase-pure germanates and applying hybrid MBE to other Ge-based oxides.

# References

1. Stapelbroek, M. & Evans, B. D. Exciton structure in the u.v.-absorption edge of tetragonal GeO$_2$. *Solid State Communications* **25**, 959-962 (1978).
2. Chae, S., Lee, J., Mengle, K. A., Heron, J. T. & Kioupakis, E. Rutile GeO$_2$: An ultrawide-band-gap semiconductor with ambipolar doping. *Applied Physics Letters* **114**, 102104 (2019).
3. Rowberg, A. J. E., Krishnaswamy, K. & Van de Walle, C. G. Prospects for high carrier mobility in the cubic germanates. *Semiconductor Science and Technology* **35**, 085030 (2020).
4. Mizoguchi, H., Kamiya, T., Matsuishi, S. & Hosono, H. A germanate transparent conductive oxide. *Nature Communications* **2**, 470 (2011).
5. Sasaki, S., Prewitt, C. T. & Liebermann, R. C. The crystal structure of CaGeO$_3$ perovskite and the crystal chemistry of the GdFeO$_3$-type perovskites. *American Mineralogist* **68**, 1189-1198 (1983).
6. Susaki, J., Akaogi, M., Akimoto, S. & Shimomura, O. Garnet-perovskite transformation in CaGeO$_3$: In-situ X-ray measurements using synchrotron radiation. *Geophysical Research Letters* **12**, 729-732 (1985).
7. Ross, N. L., Akaogi, M., Navrotsky, A., Susaki, J.-i. & McMillan, P. Phase transitions among the CaGeO$_3$ polymorphs (wollastonite, garnet, and perovskite structures): Studies by high-pressure synthesis, high-temperature calorimetry, and vibrational spectroscopy and calculation. *Journal of Geophysical Research: Solid Earth* **91**, 4685-4696 (1986).
8. Ono, S., Kikegawa, T. & Higo, Y. In situ observation of a garnet/perovskite transition in CaGeO$_3$. *Physics and Chemistry of Minerals* **38**, 735 (2011).
9. Jalan, B., Moetakef, P. & Stemmer, S. Molecular beam epitaxy of SrTiO$_3$ with a growth window. *Applied Physics Letters* **95**, 032906 (2009).
10. Zhang, H.-T., Dedon, L. R., Martin, L. W. & Engel-Herbert, R. Self-regulated growth of LaVO$_3$ thin films by hybrid molecular beam epitaxy. *Applied Physics Letters* **106**, 233102 (2015).
11. Prakash, A. *et al.* Adsorption-controlled growth and the influence of stoichiometry on electronic transport in hybrid molecular beam epitaxy-grown BaSnO$_3$ films. *Journal of Materials Chemistry C* **5**, 5730-5736 (2017).
12. Burdett, J. K., Hughbanks, T., Miller, G. J., Richardson, J. W. & Smith, J. V. Structural-electronic relationships in inorganic solids: powder neutron diffraction studies of the rutile and anatase polymorphs of titanium dioxide at 15 and 295 K. *Journal of the American Chemical Society* **109**, 3639-3646 (1987).
13. Bolzan, A. A., Fong, C., Kennedy, B. J. & Howard, C. J. Structural Studies of Rutile-Type Metal Dioxides. *Acta Crystallographica Section B* **53**, 373-380 (1997).
14. Chang, E. & Graham, E. K. The elastic constants of cassiterite SnO$_2$ and their pressure and temperature dependence. *Journal of Geophysical Research (1896-1977)* **80**, 2595-2599 (1975).
15. Wang, H. & Simmons, G. Elasticity of some mantle crystal structures: 2. Rutile GeO$_2$. *Journal of Geophysical Research (1896-1977)* **78**, 1262-1273 (1973).
16. Liferovich, R. P. & Mitchell, R. H. A structural study of ternary lanthanide orthoscandate perovskites. *Journal of Solid State Chemistry* **177**, 2188-2197 (2004).

17   Glerup, M., Knight, K. S. & Poulsen, F. W. High temperature structural phase transitions in SrSnO$_3$ perovskite. *Materials Research Bulletin* **40**, 507-520 (2005).
18   Nakatsuka, A., Arima, H., Ohtaka, O., Fujiwara, K. & Yoshiasa, A. Crystal structure of SrGeO$_3$ in the high-pressure perovskite-type phase. *Acta Crystallographica Section E* **71**, 502-504 (2015).
19   Shein, I. R., Kozhevnikov, V. L. & Ivanovskii, A. L. First-principles calculations of the elastic and electronic properties of the cubic perovskites SrMO$_3$ (M=Ti, V, Zr and Nb) in comparison with SrSnO$_3$. *Solid State Sciences* **10**, 217-225 (2008).
20   Molle, A., Bhuiyan, M. N. K., Tallarida, G. & Fanciulli, M. In situ chemical and structural investigations of the oxidation of Ge(001) substrates by atomic oxygen. *Applied Physics Letters* **89**, 083504 (2006).
21   Jones, L. *et al.* Smart Align—a new tool for robust non-rigid registration of scanning microscope data. *Advanced Structural and Chemical Imaging* **1**, 8 (2015).
22   Hohenberg, P. & Kohn, W. Inhomogeneous Electron Gas. *Physical Review* **136**, B864-B871 (1964).
23   Kohn, W. & Sham, L. J. Self-Consistent Equations Including Exchange and Correlation Effects. *Physical Review* **140**, A1133-A1138 (1965).
24   Perdew, J. P. *et al.* Restoring the Density-Gradient Expansion for Exchange in Solids and Surfaces. *Physical Review Letters* **100**, 136406 (2008).
25   Blöchl, P. E. Projector augmented-wave method. *Physical Review B* **50**, 17953-17979 (1994).
26   Kresse, G. & Joubert, D. From ultrasoft pseudopotentials to the projector augmented-wave method. *Physical Review B* **59**, 1758-1775 (1999).
27   Kresse, G. & Furthmüller, J. Efficiency of ab-initio total energy calculations for metals and semiconductors using a plane-wave basis set. *Computational Materials Science* **6**, 15-50 (1996).
28   Kresse, G. & Furthmüller, J. Efficient iterative schemes for ab initio total-energy calculations using a plane-wave basis set. *Physical Review B* **54**, 11169-11186 (1996).
29   Enslin, J. *et al.* Metamorphic Al$_{0.5}$Ga$_{0.5}$N:Si on AlN/sapphire for the growth of UVB LEDs. *Journal of Crystal Growth* **464**, 185-189 (2017).
30   Miceli, P. F. & Palmstrom, C. J. X-ray scattering from rotational disorder in epitaxial films: An unconventional mosaic crystal. *Physical Review B* **51**, 5506-5509 (1995).
31   Mountstevens, E. H., Attfield, J. P. & Redfern, S. A. T. Cation-size control of structural phase transitions in tin perovskites. *Journal of Physics: Condensed Matter* **15**, 8315-8326 (2003).
32   Wang, T., Prakash, A., Warner, E., Gladfelter, W. L. & Jalan, B. Molecular beam epitaxy growth of SnO$_2$ using a tin chemical precursor. *Journal of Vacuum Science & Technology A* **33**, 020606 (2015).
33   Jalan, B., Engel-Herbert, R., Cagnon, J. & Stemmer, S. Growth modes in metal-organic molecular beam epitaxy of TiO$_2$ on r-plane sapphire. *Journal of Vacuum Science & Technology A* **27**, 230-233 (2009).
34   Jalan, B., Engel-Herbert, R., Wright, N. J. & Stemmer, S. Growth of high-quality SrTiO$_3$ films using a hybrid molecular beam epitaxy approach. *Journal of Vacuum Science & Technology A* **27**, 461-464 (2009).

35  Moyer, J. A., Eaton, C. & Engel-Herbert, R. Highly Conductive SrVO$_3$ as a Bottom Electrode for Functional Perovskite Oxides. *Advanced Materials* **25**, 3578-3582 (2013).
36  Engel-Herbert, R., Hwang, Y., Cagnon, J. & Stemmer, S. Metal-oxide-semiconductor capacitors with ZrO$_2$ dielectrics grown on In$_{0.53}$Ga$_{0.47}$As by chemical beam deposition. *Applied Physics Letters* **95**, 062908 (2009).
37  Kajdos, A. P., Ouellette, D. G., Cain, T. A. & Stemmer, S. Two-dimensional electron gas in a modulation-doped SrTiO$_3$/Sr(Ti,Zr)O$_3$ heterostructure. *Applied Physics Letters* **103**, 082120 (2013).
38  Alcock, C. B., Itkin, V. P. & Horrigan, M. K. Vapour Pressure Equations for the Metallic Elements: 298–2500 K. *Canadian Metallurgical Quarterly* **23**, 309-313 (1984).
39  Stull, D. R. in *American Institute of Physics Handbook*   (ed Dwight E. Gray) Ch. 4k, (McGraw Hill, 1972).
40  Strephenson, R. M. & Malanowski, S. *Handbook of the Thermodynamics of Organic Compounds*.  (Elsevier, 1987).
41  Dykyj, J., Repas, M. & Svoboda, J. *Tlak Nasytenej Pary Organickych Zlucenin*.  (Vydavatelstvo Slovenskej Akademie Vied, 1984).
42  Cox, J. D. & Pilcher, G. *Thermochemistry of Organic and Organometallic Compounds*.  (Academic Press, 1970).

**Acknowledgments**

This work was primarily supported by the U.S. Department of Energy through DE-SC002021. MBE growth and characterizations (F.L, T.K.T, D.L., B.J.), were supported by the U.S. Department of Energy through DE-SC002021. F.L and T.K.T. acknowledge partial support from Air Force Office of Scientific Research (AFOSR) through Grant FA9550-21-1-0025. Parts of this work were carried out in the Characterization Facility, University of Minnesota, which receives partial support from the NSF through the MRSEC (Award Number DMR-2011401) and the NNCI (Award Number ECCS-2025124) programs. B.E.M., S.R.S., and S.A.C. carried out the STEM and XPS analysis with support from the U.S. Department of Energy, Office of Science, Division of Materials Sciences and Engineering under Award #10122 to Pacific Northwest National Laboratory (PNNL). PNNL is a multiprogram national laboratory operated for the U.S. Department of Energy (DOE) by Battelle Memorial Institute under Contract No. DE-AC05-76RL0-1830. STEM sample preparation was performed at the Environmental Molecular Sciences Laboratory (EMSL), a national scientific user facility sponsored by the DOE's Biological and Environmental Research program and located at PNNL. STEM imaging was performed in the Radiological Microscopy Suite (RMS), located in the Radiochemical Processing Laboratory (RPL) at PNNL.

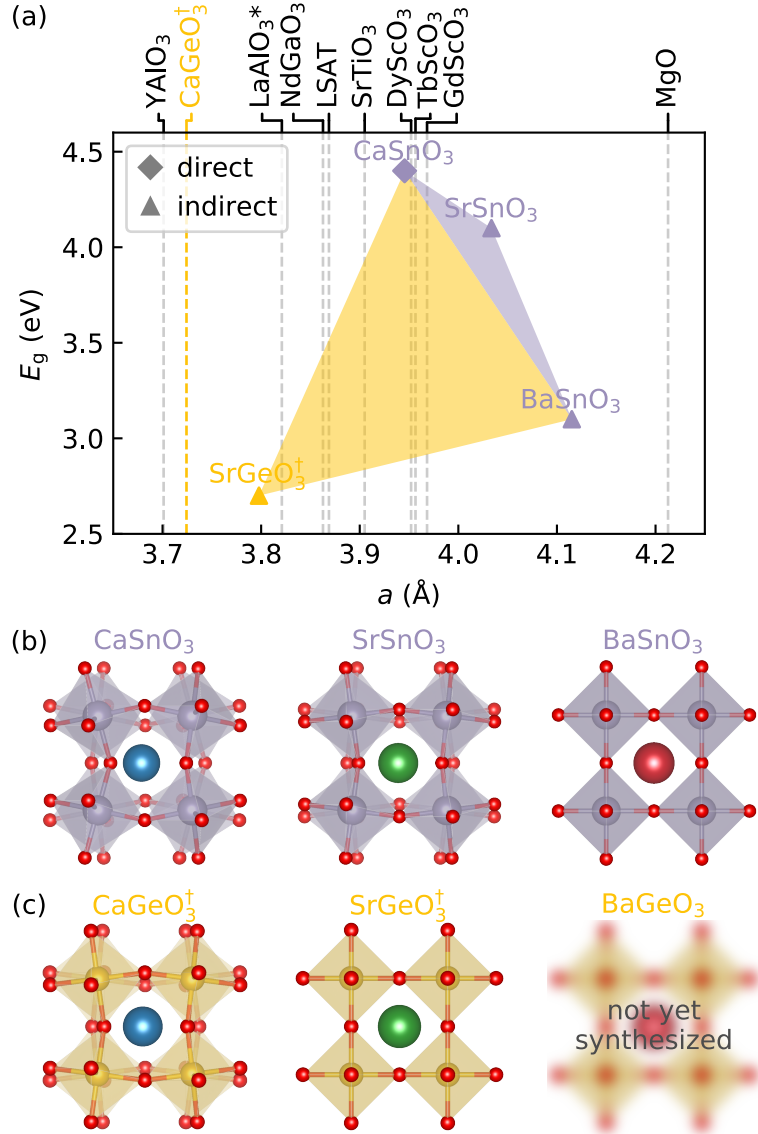

**Figure 1.** (**a**) Band gap ($E_g$) vs pseudocubic lattice parameter ($a$) of alkaline-earth stannates and alkaline-earth germanates, SrGeO$_3$. The purple shaded region represents values of $E_g$ and $a$ that are available to alloys of the three materials. The gold shaded region represents those additional values that are available when including perovskite SrGeO$_3$ in the alloy system. The lattice parameters of commercially available substrates are shown as vertical lines. (**b**) Crystal structures of alkaline-earth stannates. (**c**) Crystal structures of alkaline-earth germanates. All stannate perovskite structures are from Mountstevens and coworkers [31]. The CaGeO$_3$ structure is from Sasaki and coworkers [5]. The SrGeO$_3$ structure is from Nakatsuka and coworkers [18]. †Denotes metastable crystal structures that were quenched to ambient conditions.

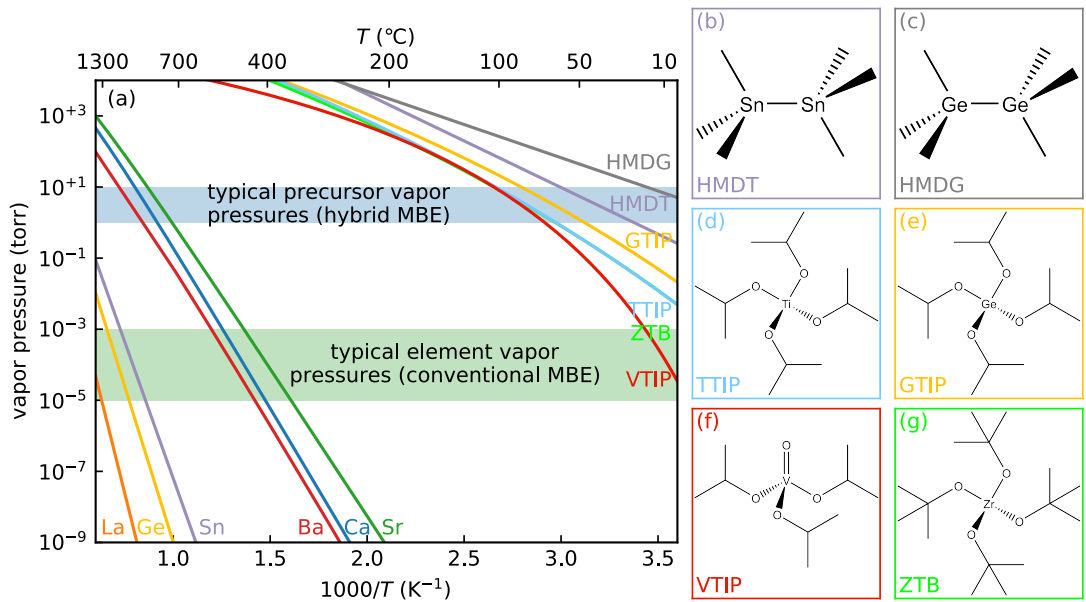

**Figure 2.** (**a**) Vapor pressures of common elements in Ge-based oxides compared to precursors for hybrid MBE. (**b-g**) The chemical structures of the precursors including hexamethyltin (HMDT, **b**) [11,32], hexamethyldigermanium (HMDG, **c**), titanium tetraisopropoxide (TTIP, **d**) [33,34], germanium tetraisopropoxide (GTIP, **e**), vanadium oxytriisopropoxide (VTIP, **f**) [35], and zirconium *tert*-butoxide (ZTB, **g**) [36,37]. All metal vapor pressures use the equations from Alcock and coworkers [38]. Ge uses a fit to data from Stull [39]. The vapor pressures for GTIP, TTIP, and VTIP use Antoine parameters from Stephenson and Malanowski [40] who themselves took this data from Dykyj and coworkers [41]. HMDT data use the enthalpy of vaporization from Cox & Pilcher [42] and the standard entropy of vaporization fit to boiling temperatures from chemical suppliers. HMDG data use Trouton's rule along with a boiling temperature provided from Sigma-Aldrich.

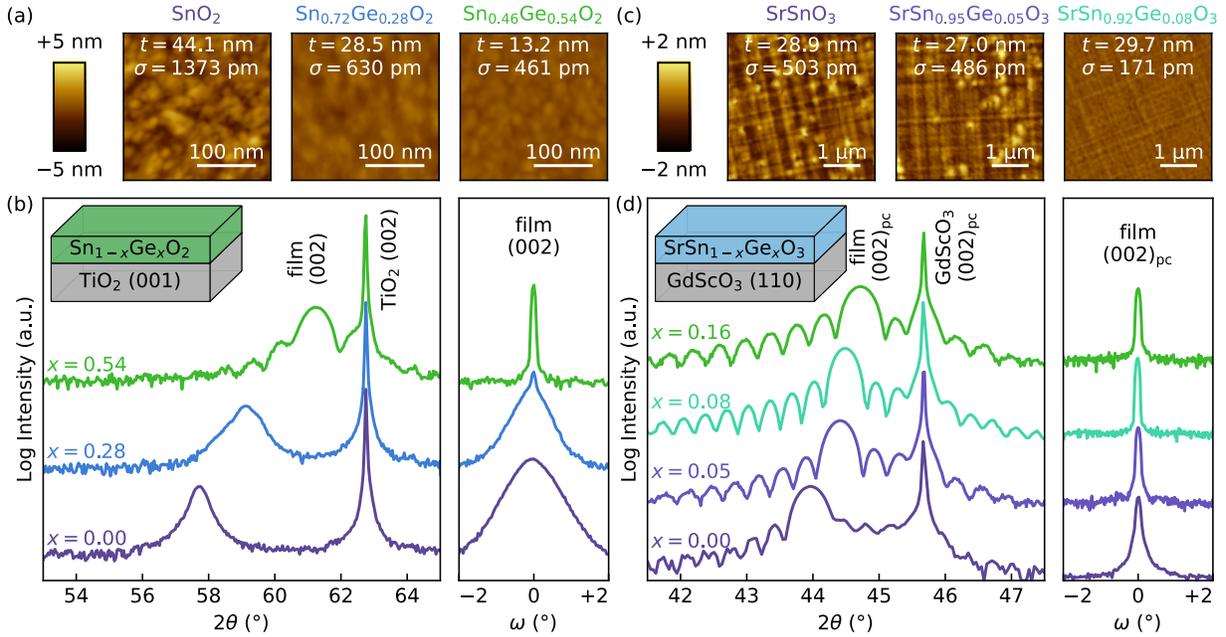

**Figure 3.** (**a**) Atomic force microscopy (AFM) of $Sn_{1-x}Ge_xO_2$/$TiO_2$ (001) films showing smooth film surfaces. (**b**) Room-temperature high-resolution X-ray diffraction (HRXRD) $2\theta$-$\omega$ coupled scans and rocking curves around the (002) film peak of $Ge_xSn_{1-x}O_2$/$TiO_2$(001) films. (**c**) AFM images of $SrGe_xSn_{1-x}O_3$/$GdScO_3$(110) showing film surfaces that get smoother with Ge incorporation. (**d**) Room-temperature HRXRD $2\theta$-$\omega$ coupled scans and rocking curves around the $(002)_{pc}$ film peak $SrGe_xSn_{1-x}O_3$/$GdScO_3$(110). The insets of (**b**) and (**d**) show the film structures.

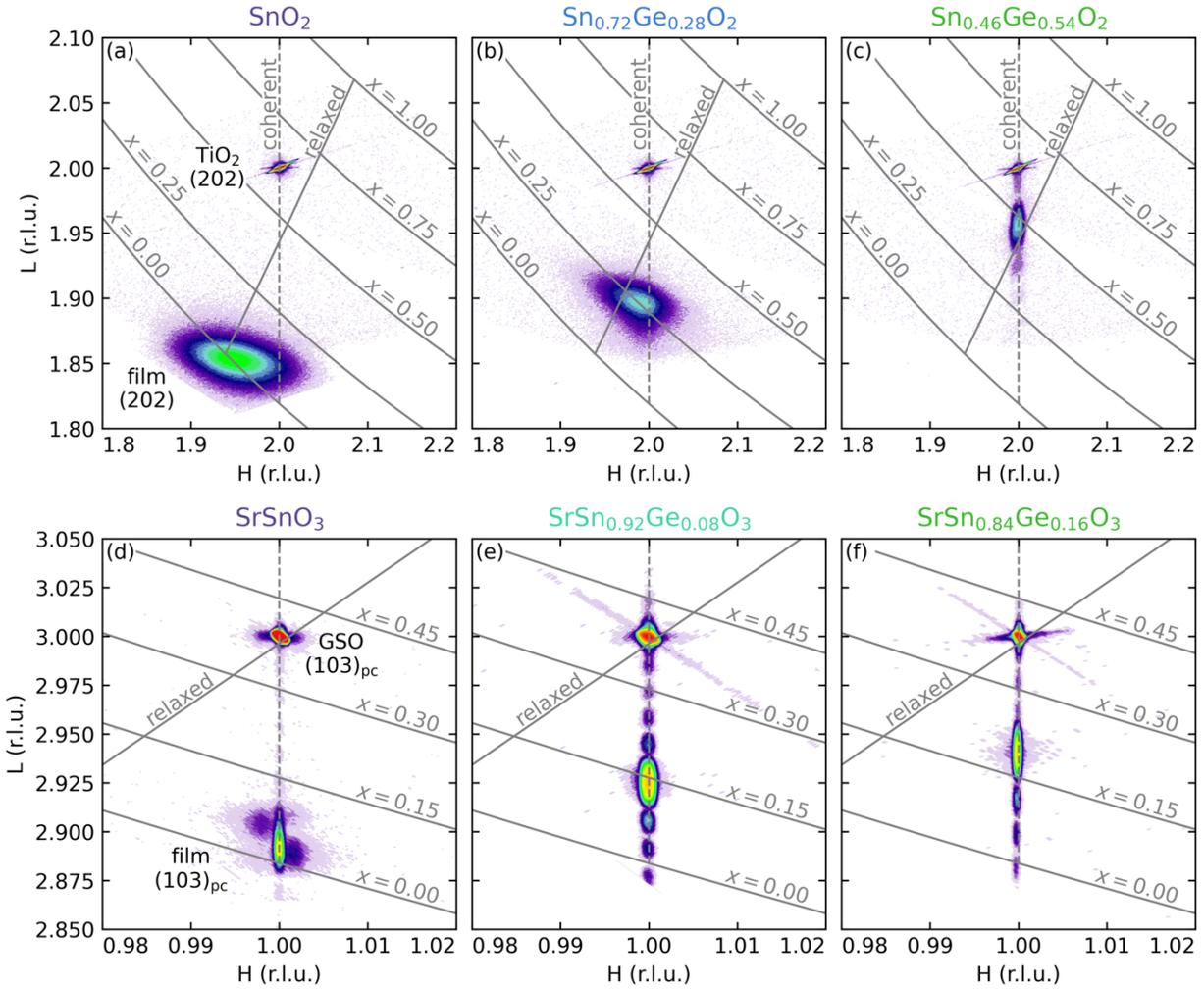

**Figure 4.** (**a-c**) Reciprocal space maps (RSMs) of Ge$_x$Sn$_{1-x}$O$_2$/TiO$_2$ (001) films of the (202) region. (**d-f**) RSMs of SrGe$_x$Sn$_{1-x}$O$_3$/GdScO$_3$ (110) films of the (103)$_{pc}$ region. All RSMs include composition contours and relaxed/coherent guidelines to show the expected peak positions based on composition and strain.

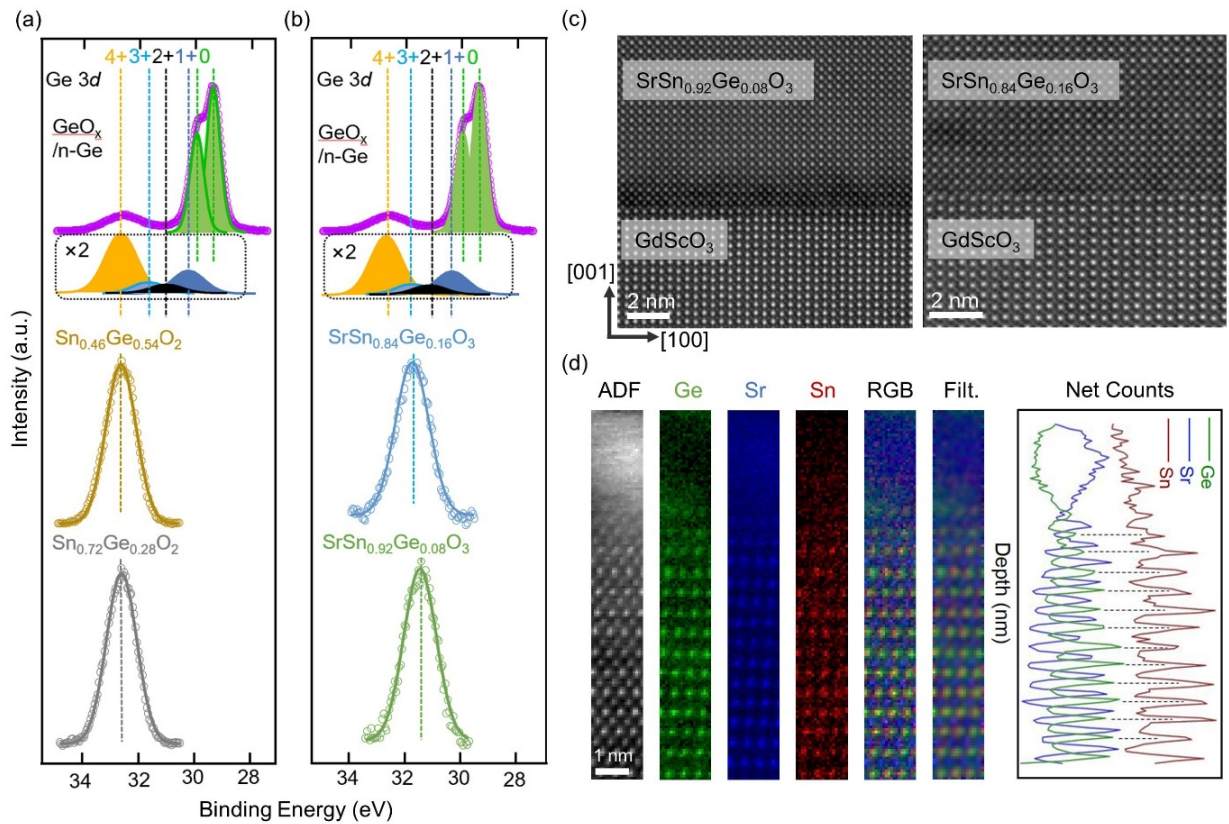

**Figure 5.** (**a-b**) Ge 3$d$ core-level hard X-ray photoelectron spectra (XPS) of $Sn_{1-x}Ge_xO_2$/$TiO_2$ (001) films (**a**) and $SrGe_xSn_{1-x}O_3$/$GdScO_3$(110) films (**b**). The top of (**a-b**) shows a $GeO_x$/n-Ge reference wafer to assist in oxidation state determination. The suboxide spectrum was vertically offset and expanded ×2. (**c**) Drift-corrected high-angle annular dark-field scanning transmission electron micrographs (STEM-HAADF) of the $SrGe_xSn_{1-x}O_3$/$GdScO_3$ interfaces. (**d**) Energy dispersive X-ray spectroscopy (STEM-EDS) of the Sn $L$ peak and electron energy loss spectroscopy (STEM-EELS) of the Sr $L$ and Ge $L$ edges. Composite maps and integrated line profiles of the $SrSn_{0.84}Ge_{0.16}O_3$ film show clear alignment of Sn and Ge signals.

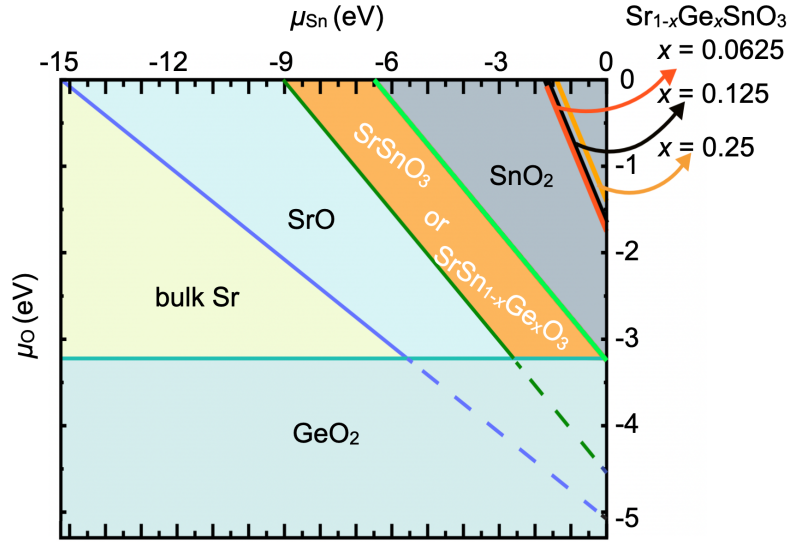

**Figure 6.** Calculated region in the tin chemical potential ($\mu_{Sn}$) vs oxygen chemical potential ($\mu_O$) plane showing where B-site alloys have lower formation enthalpy than the A-site alloys. The lines in the upper right corner ($x = 0.0625$, $x = 0.125$, and $x = 0.25$) separate the regions below which B-site alloys are preferred. This result indicates that for all allowed values of $\mu_{Sn}$ and $\mu_O$ for which SrSnO$_3$ is stable (orange region at the center), Ge will prefer to occupy the Sn site. The stability of SrSnO$_3$ is limited by the formation of SrO on the left (dark green line, corresponding to ) and the formation of SnO$_2$ on the right (light green line), i.e., $\mu_{Sn} + 2\mu_O > \Delta H^f(\text{SrSnO}_3) - \Delta H^f(\text{SrO})$, and obtain $\mu_{Sn} + 2\mu_O < \Delta H^f(\text{SnO}_2)$. The formation of GeO$_2$ poses a lower limit to the oxygen potential, as indicated in the bottom region, i.e., $\mu_{Ge} + 2\mu_O < \Delta H^f(\text{GeO}_2)$. The formation enthalpy of A-site alloy will be lower than that of B-site alloy only in the upper right corner of the $\mu_{Sn}$ vs. $\mu_O$ diagram, a region where SrSnO$_3$ itself is unstable and SnO$_2$ is favorable to form.